\documentclass{aa}
\usepackage[dvips]{graphicx}
\usepackage{psfig}
\usepackage{ltxtable}
\def\AMM{NH$_3$}
\def\NTHP{N$_{2}$H$^{+}$}
\def\HTDP{H$_{2}$D$^{+}$}  
\def\HTHP{H$_{3}^{+}$}   
\def\DTHP{D$_{3}^{+}$}   
\def\DTWHP{D$_{2}$H$^{+}$} 

\def\HCOP{\mbox{HCO$^+$}}

\def\MOLH{H$_2$}
\def\MOLN{N$_{2}$}

\def\percc{cm$^{-3}$}

\def\mic{\mbox{$\mu$m}}

\def\lesssim{\mathrel{\hbox{\rlap{\hbox{\lower4pt\hbox{$\sim$}}}\hbox{$<$}}}}
\def\gtrsim{\mathrel{\hbox{\rlap{\hbox{\lower4pt\hbox{$\sim$}}}\hbox{$>$}}}}
\begin{document}
   \title{Complete Depletion in Prestellar Cores}


  \author{C.M. Walmsley\inst{1}
\and D.R. Flower\inst{2}
\and G. Pineau des For\^{e}ts\inst{3,4}}

  \institute{INAF, Osservatorio Astrofisico di Arcetri,
             Largo Enrico Fermi 5, I50125 Firenze, Italy
         \and
            Physics Department, The University,
  Durham DH1 3LE, UK
\and
IAS, Universit\'{e} de Paris--Sud, F-92405 Orsay, France
\and LUTH, Observatoire de Paris, F-92195, Meudon Cedex, France}

   \offprints{C.M. Walmsley}

\date{June 5 version}

   \abstract{We have carried out calculations of ionization
equilibrium and deuterium fractionation for conditions
appropriate to a completely depleted,  low mass pre--protostellar
core, where
heavy elements such as C, N, and O have vanished from the gas
phase and are incorporated in ice mantles frozen on dust grain
surfaces.  We put particular emphasis on the interpretation of
recent observations of \HTDP \ towards the centre of the
prestellar core L~1544 (Caselli et al. 2003) and also
compute the ambipolar diffusion timescale.  We
consider explicitly the ortho and para forms of \MOLH,
\HTHP, and \HTDP. Our results
show that the ionization degree under such conditions depends
sensitively on the grain size distribution or, more precisely,
on the mean grain surface area per hydrogen nucleus.  Depending
upon this parameter and upon density, the major ion may be
H$^{+}$, \HTHP, or \DTHP.  We show
that the abundance of ortho-\HTDP \ observed towards L~1544
can be explained satisfactorily in terms of a complete
depletion model and that this species is, as a consequence,
an important tracer of the kinematics of prestellar cores.
   }

   \maketitle
%

\section{Introduction}
 Determining the physical structure of pre--protostellar cores is
one of the keys to understanding the development of protostars.
  In the simplest view of the development of such cores, they
contract slowly towards some ``pivotal state'', subsequent to which
dynamical collapse sets in. Clearly, one should attempt to determine
the parameters of this pivotal state, since these parameters
dictate the subsequent evolution. It seems probable that the pivotal
state is marked by high column density and low temperature. Likely
candidates can be identified using measurements of dust emission and
absorption (e.g. Andr\'{e} et al. 2000), which indicate temperatures
(of both gas and dust) of about 10~K or below, molecular hydrogen column
densities
in the range $10^{22}$ to $10^{23}$ cm$^{-2}$, and sizes of
several thousand AU.


 An additional
result from recent observational studies of prestellar
cores is that depletion of molecular species on to dust grain surfaces
is also a marker of relatively evolved cores (i.e. close to the
pivotal state) with high central densities and column densities
(e.g. Caselli et al. 2002a, Tafalla et al. 2002).
Tafalla et al.
show that CO becomes depleted by at least a factor of 10, relative to
its canonical abundance, above densities of $5\, 10^4$ \percc .
CS shows  similar behaviour,  whereas nitrogen
containing species such as \AMM \ and \NTHP \ have abundances which
are either constant or (in the case of ammonia) increase somewhat
in the high density gas surrounding the dust emission peak.
Their interpretation of these observations is that molecular nitrogen
(\MOLN), which
is the  source  of the  observed \AMM \ and \NTHP , is sufficiently
volatile to remain in the gas phase at densities of order $10^5$
\percc \ (essentially because of spot heating of the grain surfaces
by cosmic rays), whereas CO (the source of carbon for most C-containing
species) is not.

 What happens at still higher densities?
 It seems likely that, at sufficiently high densities, all species
containing heavy elements will condense out. The difference
in sublimation energies for CO and \MOLN \ is not great
(e.g. Bergin and Langer 1997) and thus one
might expect \MOLN \ and other
N--containing species to disappear from the gas phase at densities
only somewhat higher (or dust temperature slightly lower)
than found for CO. Whilst it is true that the laboratory
experiments may not accurately simulate interstellar ice surfaces,
it seems probable that \MOLN \ disappears from
the gas phase at densities above $10^6$ \percc .

 In fact, the densities inferred from mm dust continuum emission in
several objects are of order $10^6$ \percc \ and there  is
some evidence for ``holes'' in the \NTHP \ distribution at densities
above $3\, 10^5$ \percc .  For example, Bergin et al. (2002)
find a flattening of the \NTHP \ column density distribution in
B68 which could be interpreted as being due to an absence of \MOLN \
towards the dust peak. Belloche (2002) finds a ``\NTHP \ hole''
around the peak of the Class 0 source IRAM04191, which is
most easily interpreted in terms of \MOLN \ condensing out at the
highest densities in the prestellar core. Last, but not
least, Caselli et al. (2003, in what follows CvTCB)
  have detected compact emission
 (2000 AU in size) in the $1_{10}-1_{11}$ \HTDP \ line  towards the
dust emission peak of L~1544.
This latter discovery is significant in view of the fact that CvTCB
infer an abundance of order $10^{-9}$ relative to \MOLH .
Estimates of the ionization degree in L~1544  (Caselli et al. 2002a)
are of order $2\, 10^{-9}$. Taken together, these results
imply that \HTDP (and hence probably \HTHP ) is a major ion.
 It seems likely that this can be the case only if ions containing
heavy elements, such as \NTHP \ and \HCOP \, are absent, which in
turn can be true only if the heavy element content of the gas has
condensed on to grain surfaces.  If this is the case,
one must rely on species lacking heavy elements
(such as \HTDP ) to trace the kinematics of the high density
nucleus of the pre--protostellar core.  Also, it becomes relevant
to establish the ionization degree and other
physical parameters of a region where species containing heavy elements
have condensed out.

 In this paper, we study some of the consequences of complete
depletion in  an isolated, low mass prestellar core. In particular, we
compute
the ionization degree and its dependence upon physical parameters such
as the gas density and the grain size. We consider in detail deuterium
fractionation in these conditions and try to
explain the observed \HTDP \ (or, more precisely, the ortho-\HTDP ) abundance
in objects such as L~1544. In section 2, we discuss the
model assumptions and, in section 3, we summarize the results.
Then, in section 4, we consider some of the implications of
our calculations and in section 5, we briefly summarize our
conclusions. Some preliminary results of our study have been presented
elsewhere (Walmsley et al. 2004).


\section{Model}

   Our model derives from the work of Flower et al. (2003) and Flower
and Pineau des For\^{e}ts (2003).
The time--dependent chemical
rate equations are solved numerically for a situation in which heavy elements
have condensed out on to
dust grain surfaces.  As in the case of the primordial gas, the only
significant
gas--phase species  are  compounds containing
the elements H, D and He.   An important role
is played by the grains  which influence the
charge distribution (ions, charged grains, and electrons)
and  enable the formation of
\MOLH \ and HD.   Apart from these latter two processes, we neglect
surface chemistry in the present study.  We discuss below
our assumptions concerning reaction rate coefficients.

  First, we consider briefly what is implied by the statement that
``no heavy elements are present in the gas phase''.
For the purpose of the present study, this means in essence that ions
such as \HTHP , produced as a consequence of cosmic ray ionization
of molecular hydrogen, are destroyed by dissociative recombination
or recombination on grain surfaces, rather than by transferring a proton
to species such as CO and \MOLN \ which have larger proton affinities
than \MOLH .
Assuming a rate coefficient $k_{e}$ for dissociative recombination
of \HTHP \ with electrons of $4\, 10^{-7}$ cm$^{3}$ s$^{-1}$
(McCall et al. 2003, extrapolated to 10~K with a $T^{-0.52}$
dependence: see Appendix A)  and a rate coefficient for proton transfer
reactions
between \HTHP \ and species X such as CO or \MOLN \
of the order of the Langevin value,
$\alpha _{L} = 10^{-9}$ cm$^{3}$ s$^{-1}$, the condition for
``complete depletion'' is [X] $< k_{e}[\rm{e}]/\alpha_{L}$, where
the square brackets denote a fractional abundance; this implies that,
for example, the CO abundance [CO] in the ``complete depletion
limit'' should be less than 400[e],  or
roughly $4\, 10^{-6}$ in cores such as L~1544,
where [e] $\approx\ 10^{-8}$.
Thus, the calculations reported here will be relevant when species
such as CO have abundances much less than
$10^{-6}$.

 In what follows, we report the steady--state conditions predicted by our
calculations. The timescales to reach chemical equilibrium
are somewhat shorter than the dynamical timescales in cores such as
L~1544. The free--fall time is approximately $3\, 10^4$ y for a density
of $10^6$ cm$^{-3}$;
but, when magnetic fields are present, the collapse time is about an
order of magnitude larger (see Ciolek and Basu 2000, Aikawa et al. 2003).
As shown below, the equilibrium timescale for the entire
chemistry is
roughly $10^5$ y, although ionization equilibrium (with a characteristic
time $1/[k_{e}n(e)]$, which is of the order of a few years)
is reached much more rapidly.
We conclude that the steady--state assumption is a valid
first approximation but that some of our results may need
to be reconsidered in the light of their time dependence.

\subsection{Grain processes}
  In general, we have followed the approach and adopted the 
rate coefficients
   relating to ``large grains'' of Flower and
Pineau des For\^{e}ts (2003, in what follows FPdesF;  see their
Appendix A).
 The following processes
affecting the grain charge are taken into account: electron attachment;
electron detachment by the H$_2$ fluorescence photons, generated by
collisions with
the secondary
electrons produced by cosmic ray ionization of hydrogen; attachment of
positive ions to neutral grains and recombination of positive ions with
negatively charged grains.
However, for the present applications,
their assumption of a Mathis et al. (1977, hereafter MRN)
power law distribution of grain sizes
(based on the extinction curve derived for diffuse interstellar gas) seemed
inappropriate. We have therefore assumed single--size grains,
with a dust--to--gas mass density ratio  (0.013)
consistent with all elements heavier than helium having
condensed out.  Moreover, we adopt a grain material density
of 2 gm cm$^{-3}$ for the core plus ``dirty ice'' mantle.
We treat the grain radius $a_{g}$ as a parameter of the model.

Recombination of ions with electrons on negatively charged grain surfaces is
important because, if the available grain
surface area is large enough, this process can dominate the destruction
of ions such as \HTHP .  Recombination on grains is certainly the main
destruction
mechanism for protons, which recombine only slowly (radiatively)
with electrons in the gas phase, and  hence  our results depend
on the surface area of negatively charged grains.
Accordingly, we have considered carefully the
distribution of grain charge, adopting the rate coefficients of FPdesF.
We assumed that \MOLH \ and HD form on grain surfaces, at
rates which scale with the total grain surface area.

\subsection{Chemical scheme and D-fractionation}
The assumption of  complete depletion of the heavy elements
on to grains leads to a considerable reduction in the number
of chemical species which have to be taken
into account: H, \MOLH, H$^{+}$, H$_{2}^{+}$, \HTHP\ (and their
deuterated forms), He and He$^{+}$. At
the low temperatures which prevail in protostellar cores,
deuterium fractionation occurs, resulting in greatly enhanced
abundances of \HTDP , \DTWHP\ and \DTHP .
In this situation, the rates of the associated forwards
and reverse deuterium--substitution reactions depend not only on
the differences in
zero--point energies along the deuteration sequence
(Ramanlal et al. 2003) but also on the ortho/para abundance
ratios of \MOLH , H$_{2}^{+}$, \HTHP , and \HTDP .
Accordingly, we have treated the ortho and para forms
of these molecules as separate species in our calculations.

A key role is
played by the ortho/para ratio of molecular hydrogen, which is
expected to be much higher than in
LTE (see the discussion of Le Bourlot 1991); this has
consequences for the ortho/para ratios of species such as \HTDP \
(see Gerlich et al. 2002, hereafter GHR; Pagani et al. 1992),
which become enhanced
roughly in proportion to that of \MOLH .
Also important in this context are the spin
selection rules for ortho--para conversion, discussed, for example,
by Uy et al. (1997).

Caution has to be exercised when using the data of Uy et al. (1997). These
authors considered conditions in which the ortho and para forms are
statistically populated, i.e. in proportion to their nuclear spin
statistical weights, (2$I$ + 1). This situation obtains at temperatures and
densities much higher than those considered here. In prestellar cores, only
the low--lying states of the ortho and para species are significantly
populated. Under these circumstances, the statistical weights of the
rotational levels, (2$J$ + 1), must be taken explicitly into account. When
compiling the rate coefficients for reactions involving the ortho and para
forms of \MOLH , H$_{2}^{+}$, \HTHP , and \HTDP , we made the simplifying
assumption that only the lowest rotational level is populated in each case
and treated the corresponding ortho and para forms as distinct species,
which interconvert through chemical reactions. The complete set of
reactions included in the model and their rate coefficients are given in
Appendix A.

In the limit
of complete depletion, deuterium fractionation
occurs mainly in the reactions

\begin{equation}
{\rm H}_{3}^{+} \  + {\rm HD} \ \rightarrow \
 {\rm H}_{2}{\rm D}^{+} \  + {\rm H}_{2}
\label{eqd1}
\end{equation}
with rate coefficient $k_{1}$,

\begin{equation}
 {\rm H}_{2}{\rm D}^{+} \ + {\rm HD} \ \rightarrow \
 {\rm D}_{2}{\rm H}^{+} \ + \ {\rm H}_{2}
\label{eqd2}
\end{equation}
 with rate coefficient $k_{2}$, and

\begin{equation}
{\rm D}_{2}{\rm H}^{+} \ + \ {\rm HD} \rightarrow \
{\rm D}_{3}^{+} \ + \ {\rm H}_{2}
\label{eqd3}
\end{equation}
 with rate coefficient $k_{3}$.
These reactions are exothermic by roughly 200~K
and consequently
enhance abundance ratios such as
$n({\rm H}_{2}{\rm D}^{+})/n({\rm H}_{3}^{+})$.  However, such enhancement
can be
limited by a variety of processes.
If heavy elements are present in the gas phase,
the high abundances of molecules with high proton affinities, such as CO,
result in
the fractionation being transferred to ions such as DCO$^{+}$;
  this  limits the degree of enhancement of deuterium in
\HTDP .  It follows that the
depletion of heavy elements favours deuterium fractionation (see
Dalgarno and Lepp 1984), and so a ``completely depleted core'' is a
limiting case
where one might expect high degrees of enhancement.

 Even in the
absence of heavy elements, there are several processes which
limit deuterium substitution in  species such as \HTHP .
For example, it has been known for a long time
that  dissociative recombination with electrons,
at a rate $k_{e}\,n(\rm{e})$,
competes with deuterium enrichment in \HTDP , for example
 (Watson 1978). It is also clear that recombination of \HTDP , \DTWHP\
 and \DTHP\
on grain surfaces can have a similar effect, and so models
with a large grain surface area tend to have moderate
D-fractionation.  The relative importance of these two processes is
roughly the same for D-fractionation and for ion neutralization:
in models where the D-fractionation is limited
by recombination on grain surfaces, the degree of ionization
is limited by the same process.

 D-fractionation in an ion such as \HTDP \ is determined also by
the transfer of the deuterium to the higher members of the
chain, \DTWHP \ and  \DTHP . Thus reactions~\ref{eqd2}
and~\ref{eqd3} restrict the fractionation in
\HTDP \ and \DTWHP , respectively. As we shall see below, the
effect of these reactions is to limit ratios such as
[\HTDP]/[\HTHP] and [\DTWHP]/[\HTDP] to values of the order of
unity.

GHR pointed out that
ortho-\MOLH \ can   restrict
deuterium fractionation  because
the reaction between ortho-\MOLH \ and ortho-\HTDP \ is exothermic and
believed to be
fast. Depending on internal excitation, this process
may also be important for \DTWHP .
We find that the reaction with ortho-\MOLH \
is not usually dominant but is
significant because the abundance of ortho-\MOLH \ is high.

The ratio of the \HTDP \ and \HTHP \ abundances
is given approximately by

\begin{equation}
[{\rm H_2D^+}]/[{\rm H}_3^+] = \frac{2 \, k_1[{\rm D}]}{k_e[{\rm
e}] + k_{ig}[{\rm g}^-] + 2k_2[{\rm D}] + 2k^-_1}
\label{equil}
\end{equation}
 where [D] is the fractional abundance of D in the gas phase,
 $k_{ig}$ is the rate coefficient for neutralization
of \HTDP\ on negatively charged grains of number density
$n_{{\rm g}^{-}}=[{\rm g}^{-}]\, n_{\rm H}$, and $k_{1}^{-}$ is
the rate coefficient for the reverse of reaction~\ref{eqd1}.
We see that
deuterium fractionation is limited by the
terms in the denominator of the right hand side of this equation.   One may
note
also that, whilst [\HTDP]/[\HTHP] and  [\DTWHP]/[\HTDP] are limited
to $k_{1}/k_{2}$ and $k_{2}/k_{3}$, respectively, there is no
such restriction for [\DTHP]/[\DTWHP] because
 \DTHP \ does not react with
HD.  Thus,  in certain circumstances, \DTHP \ can be
the principal ion.

\subsection{Numerical method}

 The code described by Flower et al. (2003)
has been used to calculate the steady--state abundances of the
chemical species, X$_{i}$.
The program integrates the coupled differential equations
\begin{displaymath}
\frac{d}{dt}\, n({\rm X}_{i})\, = \, f[n({\rm X}_{i}),T,t]
\end{displaymath}
 for the number densities $n({\rm X}_{i})$, as functions of time,
 $t$, until steady state is attained.
The kinetic temperature, $T$, was
assumed constant. By adopting this
approach, not only are the steady--state values of the
number densities obtained for sufficiently large $t$
but  also information is recovered on the timescales
required for the various chemical reactions to reach equilibrium.

\begin{figure}
\centering
\resizebox{\hsize}{!}{\includegraphics[height=8cm]{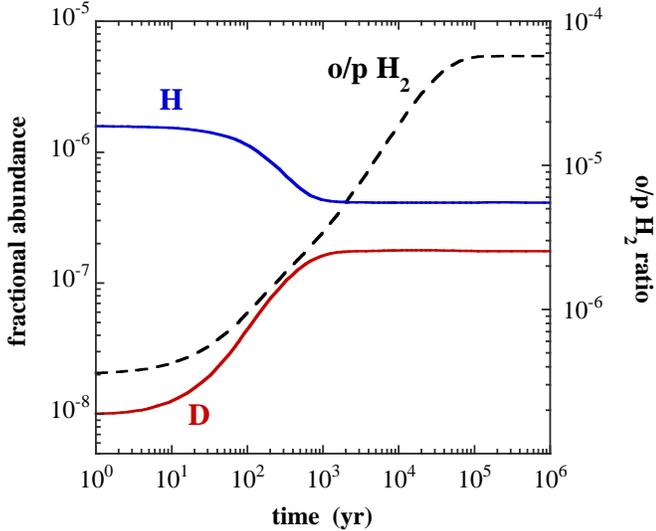}}
  \caption
  { Sample result for our reference model ($n(\rm{H}_2) = 10^6$ \percc ,
  $T$ = 10~K, $a_g$ = 0.1 \mic ,
$\zeta =\, 3\, 10^{-17}$ s$^{-1}$) showing the approach
to steady--state conditions for
ortho/para \MOLH , D and H. }
  \label{time_dep}
\end{figure}

 In Fig.~\ref{time_dep}, we show  the results of a sample
calculation  in which, initially, the hydrogen and deuterium
are in molecular form, with an elemental abundance ratio
$n_{\rm D}/n_{\rm H} = 1.6 \times 10^{-5}$. 
The initial ortho/para \MOLH \
ratio was $3.5 \times 10^{-7}$, the equilibrium value at the gas kinetic
temperature $T = 10$~K (assumed constant).
 We find that the time required to reach steady state
for all species is of
the order of $10^5$ y (almost independent of number
density). However, as noted above, the
time to attain
ionization equilibrium (a few years) is much shorter than that required to
attain
equilibrium between ortho- and para-\MOLH \ (the longest chemical timescale).
Atomic
hydrogen and deuterium are intermediate,
reflecting the timescale for the formation
of \MOLH \ and HD on grains.

 The time characterizing chemical steady state
is comparable with the
free--fall time, $\tau_{\rm ff}$, at the lowest density
($n_{\rm H}$ = $2\times 10^5$ \percc) and is an order of magnitude
larger than $\tau_{\rm ff}$ at the highest density
($n_{\rm H}$ = $2\times 10^7$ \percc ) considered.
However, magnetic support to the cloud increases the
time for core formation and, in practice, the timescale for
chemical equilibrium is likely to be comparable with that for
contraction and collapse of
the protostellar core.

\section{Results}
\subsection{Dependence on model parameters}
 We present our results with reference to a
model with the following parameters :
$a_{g}\,=\,0.1$ $\mic$, $\rho_{g}\,=\,2$ g \percc, $T$ = 10~K, and
$\zeta\,=\, 3\times 10^{-17}$ s$^{-1}$, where $a_{g}$ is the
grain radius, $\rho_{g}$ is the mean density of the grain
material (core and ice mantle), $T$ is the gas temperature,
and $\zeta$ is the rate of cosmic ray ionisation of \MOLH .
  To these parameters
correspond a fractional abundance of grains $n_{g}/n_{\rm H}$
= $3.5\times 10^{-12}$ and a grain surface area per hydrogen
nucleus $n_{g}\, \sigma _{g}/n_{\rm H}\, =\, 1.1\, 10^{-21}$
cm$^{2}$. These numbers
have been derived assuming that all heavy elements are
incorporated into either the cores or the ice mantles of the
grains (corresponding to a dust--to--gas mass density ratio
of 0.013).

 \begin{figure}
\centering
\resizebox{\hsize}{!}{\includegraphics[height=8cm]{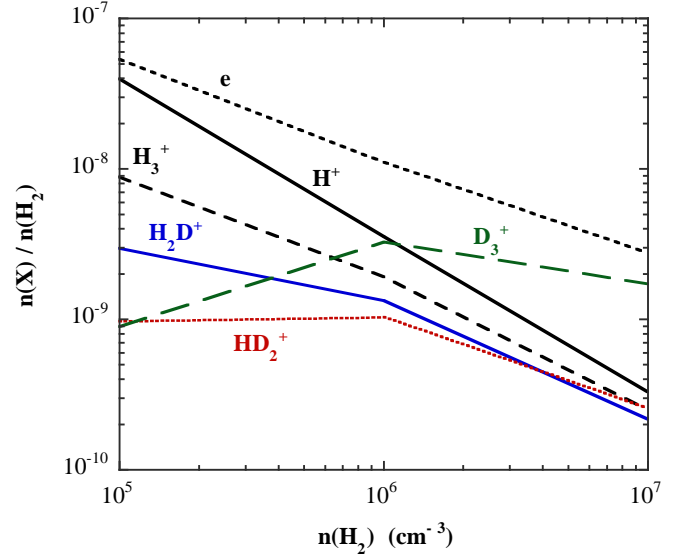}}
\caption{Abundances of major ions and electrons in our
standard model ($a_g$ = 0.1 \mic , $T$ = 10~K,
$\zeta \,=\, 3\times 10^{-17} s^{-1}$)
as functions of the density of
molecular hydrogen.}
 \label{standard}
 \end{figure}

\begin{figure}

\centering
\resizebox{\hsize}{!}{\includegraphics[height=8cm]{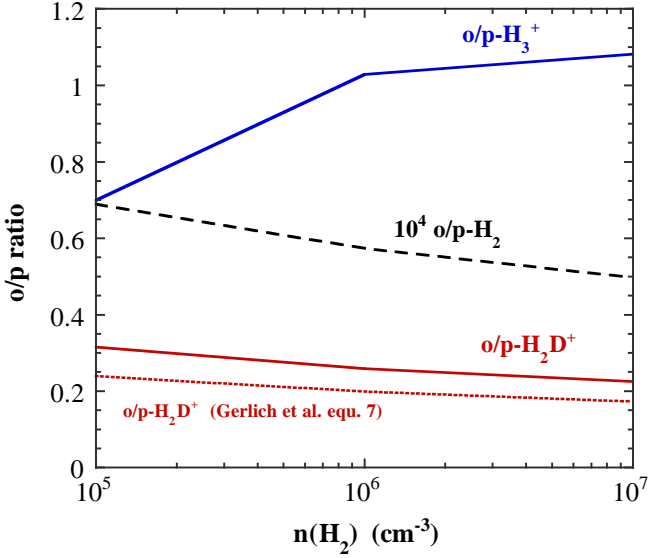}}
\caption{Ortho/para ratios of \MOLH\ (multiplied by $10^4$),
\HTHP , and \HTDP  as functions of
density for our reference model (cf. Fig.~\ref{standard}).
 The approximate expression of Gerlich et al. (2002) for \HTDP \ is  shown
for comparison.    }
\label{ortpara}
\end{figure}

 In Fig.~\ref{standard} are plotted the fractional abundances of
the principal ions and of the electrons as functions of
$n({\rm H_2})\approx\,n_{\rm H}/2$; the results of computations of
ortho/para ratios for the same model are presented in Fig.~\ref{ortpara}
and will be discussed in section 3.4 below.
 At the lowest density considered, $n$(\MOLH) =
$10^5$ \percc , H$^{+}$ is the major ion and the removal of
molecular ions by dissociative recombination is significant.
H$^{+}$, on the other hand, is removed by recombination on grain
surfaces.  As $n$(\MOLH) increases, recombination on grains
dominates dissociative recombination for the molecular ions.
H$^{+}$ becomes a relatively minor species at high
density.

The balance between the various deuterated forms of \HTHP \
is determined approximately by
equation~\ref{equil} and its analogues for [\DTWHP]/[\HTDP]
and [\DTHP]/[\DTWHP].  At densities below $10^6$
\percc , the dissociative recombination term
$k_{e}\, $[e] is dominant in the denominator
of the right hand side of equation~\ref{equil}, whereas,
at higher densities, the term due to the reaction with HD
becomes more important and results in [\HTDP]/[\HTHP]
(and [\DTWHP]/[\HTDP]) being close to unity.
On the other hand, \DTHP\ cannot be destroyed by HD and becomes
the dominant ion.

\begin{figure}
\centering
\resizebox{\hsize}{!}{\includegraphics[height=8cm]{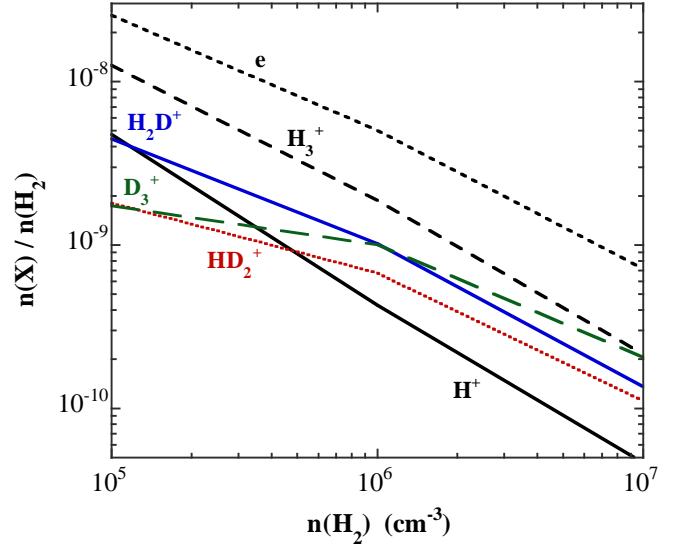}}
\caption{ Abundances of major ions and electrons in a
model with a smaller grain size of 0.025 \mic .
    }
\label{abund025}
\end{figure}
 Fig.~\ref{abund025} is analogous
to Fig.~\ref{standard}, for
an assumed grain radius of 0.025 \mic . In this case, recombination
of ions on grain surfaces is the dominant neutralization process for
all ions and H$^{+}$ is never dominant. Moreover, for
densities below $10^7$ \percc, \HTHP \ is more abundant than its deuterated
forms.  We note that the  increased rate of grain
neutralization causes the ion abundance to fall and its density
dependence to become steeper.

\subsection{Ionization degree for complete depletion}
   The determination of the ionization degree in the region of
complete depletion is important for a number of reasons.
The degree of ionization determines the coupling of the magnetic
field to the gas and hence the timescale for
ambipolar diffusion. The electron abundance
plays a crucial role also in the chemistry
and in deuterium fractionation.

\begin{figure}
\centering
\resizebox{\hsize}{!}{\includegraphics[height=8cm]{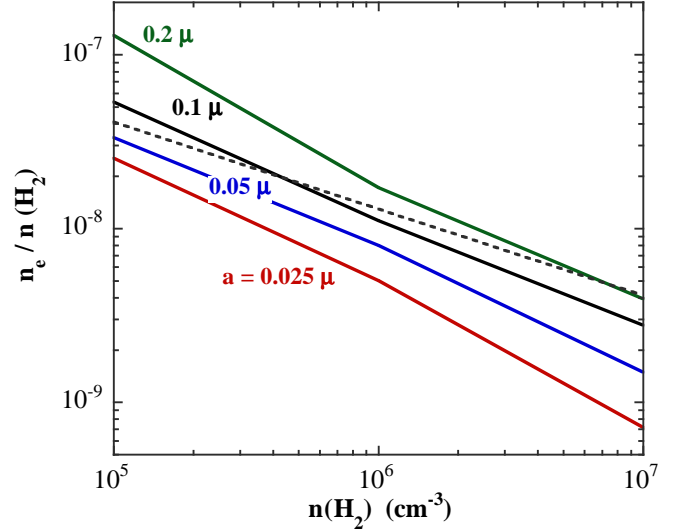}}
\caption{The  free electron abundance computed as a
function of density for grain sizes between
0.025 and 0.2 \mic . The result of McKee (1989)
is shown for comparison (broken curve).  Note the decrease in the level of
ionization for small grain sizes (large grain surface areas).    }
\label{fig_xe}
\end{figure}

 In Fig.~\ref{fig_xe}, the  free electron abundance,
 $n$(e)/$n(\rm{H}_2)$, computed as a function of
density, is compared with the relationship of McKee (1989), $n({\rm
e})/n(H_2) = 1.3\times 10^{-5}n(H_2)^{-0.5}$,
which is often used
in discussions of core evolution. It may be seen that our model generally
predicts lower degrees of ionization. We find that $n$(e)/$n(\rm{H}_2)$
varies approximately as $n_{\rm H}^{-0.75}$
between $n_{\rm H}\, =\,2\times 10^6$ and $2\times 10^7$
\percc .
As already noted, the  degree of ionization of the gas is sensitive
to the assumed grain size, owing to the importance of the recombination
of ions with electrons on grain surfaces. The fraction
of electrons attached to
grains remains small for all the grain sizes considered here:
the ratio of negatively charged grains to free electrons attains 0.20 for
for $a_g$ = 0.025 \mic \ and $n({\rm H}_{2}) = 10^7$ \percc . However, we
note that this ratio becomes larger for smaller grain sizes, becoming $>>
1$ for $a_g < 0.01$ \mic . The fraction of grains which is neutral
decreases from 0.68 for $a_g$ = 0.025 \mic \ to 0.20 for
$a_g$ = 0.2 \mic \ and for $n({\rm H}_{2}) = 10^7$ \percc \ (but this ratio is
insensitive to the gas density).
 The fraction of positively charged grains
is always negligible.

\subsection{Ambipolar Diffusion in depleted cores}
 If the core is
sub--critical (i.e. if the magnetic field is capable of resisting
collapse), the timescale for star formation is determined
by that for ambipolar diffusion.  Here we estimate the timescale for
ambipolar diffusion under the conditions of complete heavy element depletion.
%
\begin{figure}
\centering
\resizebox{\hsize}{!}{\includegraphics[height=8cm]{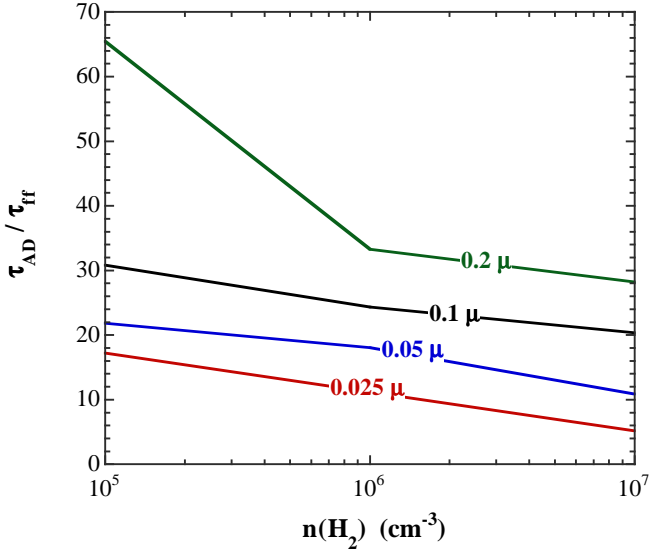}}
\caption{Ratio of ambipolar diffusion time
$\tau _{\rm AD}$ (equation~\ref{ambip}) to
free fall time $\tau_{\rm ff}$ as function of molecular
hydrogen density for values of the grain radius between
0.025 and 0.2 \mic . }
\label{ambi}       
\end{figure}

The ambipolar diffusion timescale, $\tau _{\rm AD}$, depends
on the ion abundance, $n_i/n_n$,
the rate coefficient for momentum transfer between the ions and
the neutrals (essentially \MOLH), $<\sigma v>_{in} \approx 2\times 10^{-9}$
cm$^3$ s$^{-1}$, and the masses $m_i$ and $m_n$ of the charged and neutral
collision partners.  We have taken

\begin{equation}
\tau_{\rm AD}\, \approx\, \frac{2}{\pi G\, m_{n}^2}
\sum _{i} \frac{n_{i}}{n_{n}}
\frac{m_{i}m_{n}}{m_{i}+m_{n}}<\sigma v>_{in}
\label{ambip}
\end{equation}
where $G$ is the gravitational constant
and the summation is over the
ionic species.  The rate coefficient for momentum transfer
approaches its Langevin form

\begin{displaymath}
<\sigma v>_{in}=2\pi e(\alpha /m_{in})^{0.5}
\end{displaymath}
at low temperatures (cf. Flower 2000), and hence
$\tau_{\rm AD}$ becomes proportional to the square root of the reduced mass,
$m_{in} = [m_{i}m_{n}/(m_{i}+m_{n})]$;
$\alpha = 7.7\times 10^{-25}$ cm$^3$ is the polarizability of H$_2$.
Thus, $\tau_{\rm AD}$ increases by 50 percent as the dominant ion changes
from being H$^{+}$ to being D$_{3}^{+}$.
In Fig.~\ref{ambi}, we plot the ratio of the
ambipolar diffusion timescale
$\tau_{\rm AD}$ to the free--fall timescale $\tau_{ff}$
as a function of density. We see from Fig.~\ref{ambi} that the
ambipolar diffusion timescale in the limit of complete depletion is
of the same order of magnitude as that inferred for
normal ionic abundances in dense cores, i.e. about an order of magnitude
larger than the free--fall time (see, for example, Shu et
al. 1987, Ciolek and Basu 2000).  As one might expect, smaller grain sizes
lead to shorter ambipolar diffusion timescales, owing to more rapid
neutralization on grain surfaces.

The above discussion neglects the collisional coupling between the neutral
gas and charged grains (cf. Nakano and Umebayashi 1980). For the conditions
that we have considered, the coupling to the charged grains is dominant,
particularly for small grain sizes and high gas densities. We estimate that
its effect is to increase the value of $\tau_{AD}$ by one to two orders of
magnitude.
We conclude that the collapse of a magnetically sub--critical,
completely depleted core is likely to be impossible on reasonable
timescales.

\subsection{The abundances of ortho-\MOLH ,
ortho-\HTDP and ortho-\HTHP}

Fig.~\ref{ortpara} shows that the ortho/para ratios
considered here are relatively insensitive to density.
For molecular hydrogen, we derive a
steady--state ortho/para
ratio of approximately $5\, 10^{-5}$
for the conditions in the nucleus of
L~1544, and this value is reflected in
the \HTDP \  ortho/para ratio, as shown by GHR.
In fact, we find that the \HTDP \ ortho/para ratio
predicted by equation (7) of GHR is
a good approximation to our numerical
results.

 The expression of
Le Bourlot  (1991) for ortho/para \MOLH \ was derived
for different conditions (lower density and high abundances of
molecular ions such as \HCOP ) and is not a good approximation
in the conditions that we have considered.
In our situation, the ortho/para \MOLH \ ratio is given approximately
by $\zeta /[k_{op}\,n(\rm{XH}^{+})]$,
where $n(\rm{XH}^+)$ represents the sum of the number densities of ions such as
H$^{+}$ and \HTHP \ which convert ortho- to para-\MOLH \ through proton
transfer reactions, with a rate coefficient for ortho to para conversion
$k_{op}$ which is of order $10^{-10}$
cm$^{-3}$ s$^{-1}$.

\begin{figure}
\centering
\resizebox{\hsize}{!}{\includegraphics[height=8cm]{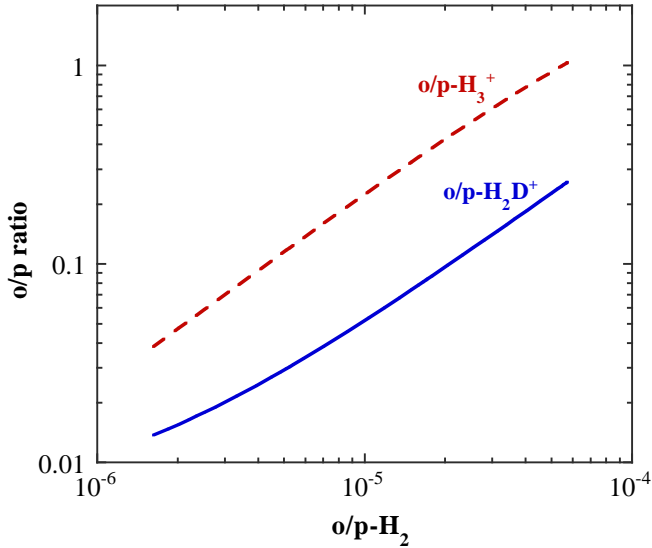}}
\caption{ Ortho/para ratios of \HTHP \ and
\HTDP \ as
functions of the molecular hydrogen ortho/para ratio.
The conditions are those of our reference model:
$n$(\MOLH)= $10^6$ \percc , $T$ = 10~K,
$a_{g}$ = 0.1 \mic , $\zeta = 3\times
10^{-17}$ s$^{-1}$. }
\label{fig_orpara}
\end{figure}

The ortho/para
\HTDP \ ratio has a flat dependence on density,
reflecting the behaviour of ortho/para \MOLH .  The ortho forms of both
\HTDP \ and \HTHP \
are produced mainly in reactions of the para forms with ortho-\MOLH \, and
so their high ortho/para ratios are attributable to the relatively high
ortho-\MOLH \ abundance; Fig.~\ref{fig_orpara} illustrates
this point. Because the ortho/para \MOLH \ ratio is not thermalized
under the conditions
which we consider, the ortho/para \HTDP \ and \HTHP \ ratios are not
thermalized  either. It follows that measurements of the
ortho/para \HTHP \ ratio should {\it not} be used to infer a temperature.

\begin{figure}
\centering
\resizebox{\hsize}{!}{\includegraphics[height=8cm]{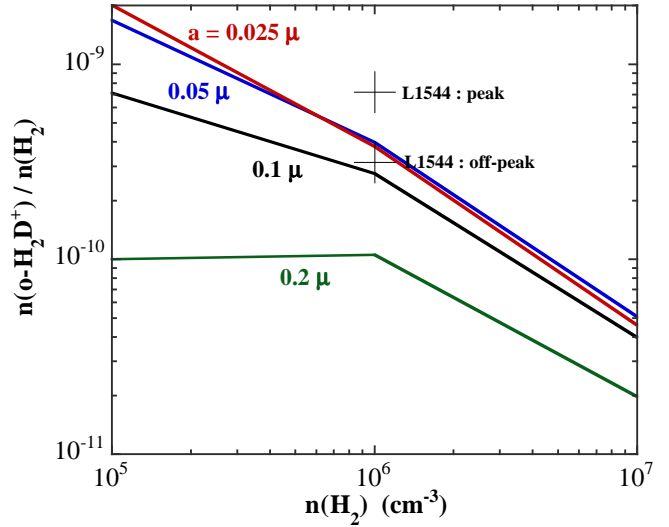}}
\caption
{ The computed ortho-\HTDP \ abundance from our reference
model ($T$ = 10~K and
$\zeta \, = \, 3\times \, 10^{-17}$ s$^{-1}$)
as a function of density for several values of the
assumed grain radius.
The values observed by Caselli et al. (2003) are shown for
comparison. }
\label{orh2dab}
\end{figure}

The basic question that we wish to answer
is whether the ortho-\HTDP
\ emission observed in L~1544 can be reproduced by a model
which assumes complete heavy element depletion.
 The abundance of ortho-\HTDP \,
relative to \MOLH \, observed towards the peak of  L~1544 is
$5\, 10^{-10}$ according to CvTCB.
In Fig.~\ref{orh2dab} are shown the computed values of
[ortho-\HTDP ]/[\MOLH ] as a function of density for various assumed grain
sizes.  There is a tendency for [ortho-\HTDP ]/[\MOLH ] to decrease
with increasing density, owing to the fact that deuteration proceeds
further down the chain (towards \DTHP )
at high density.  Our results are within a factor of
2 of the observed peak ortho-\HTDP \ abundance, for grain radii
below 0.1 \mic .
Given the observational uncertainties, this level of agreement might
be considered satisfactory.  It is also significant that computations
for $a = 0.01$ \mic \ show that the computed ortho-\HTDP \ abundance
decreases for all values of
$n$(H$_2$)  relative to the results in Fig.~\ref{orh2dab} for
$a = 0.025$ \mic , and thus
grain radii in the range 0.025--0.05 \mic \ maximize the
ortho-\HTDP \ abundance.
We note that the ``observed'' abundance
assumes a rate of collisional excitation of
\HTDP , a temperature, and a density distribution, of which
the latter two are controversial (see, for example, Galli et al. 2002).
Higher angular resolution studies of the mm--submm continuum
are needed for further progress.

One result from our study is that a
relatively high rate of cosmic ray ionization is required in the nucleus of
L~1544 to account for the observed ortho-\HTDP \ emission;
 this implies the penetration of low  energy (100 MeV)
cosmic rays into the high density core.
 Cosmic ray fluxes higher than assumed here would enhance
the computed ortho-\HTDP \ abundance; but to a higher flux would correspond
an increased rate of heat input to the gas
and temperatures in excess of those observed in \AMM \
and other tracers.

\subsection{Atomic Deuterium in the completely depleted core}
\begin{figure}
\centering
\resizebox{\hsize}{!}{\includegraphics[height=8cm]{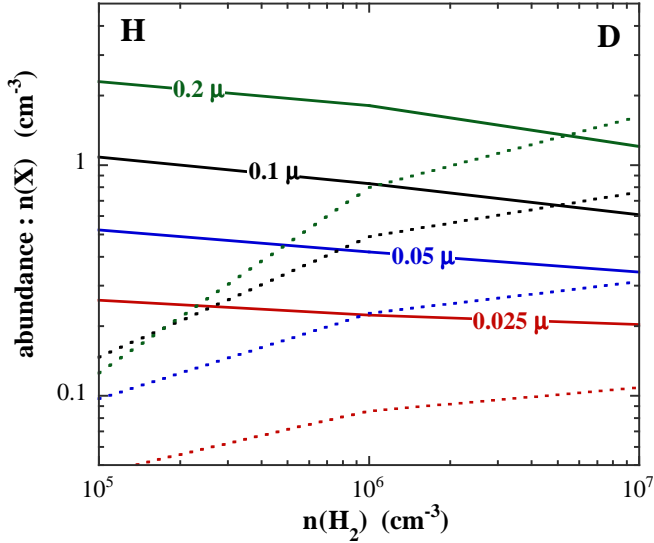}}
\caption{ Atomic H (full lines) and atomic D (broken lines) densities for
four values  of the
grain radius (increasing from bottom to top) for densities
between $10^5$ and $10^7$ \percc .
    }
\label{atomicd}
\end{figure}

 An interesting by--product of our work is the prediction of
the atomic D abundance under the conditions of complete depletion.
Fig.~\ref{atomicd} shows the abundances of atomic deuterium
and hydrogen as functions of density for various grain radii.
Atomic H is produced
mainly by cosmic ray ionization and destroyed by the formation of
\MOLH \ on grains. As a result, the fractional
abundance of H varies inversely with the density and decreases
with the grain radius, owing to the greater grain surface area
available to form \MOLH .  Atomic D, on the other hand, is produced
by the recombination of deuterated ions and becomes relatively more
abundant as deuterium fractionation increases.  It is
noteworthy that the
abundance of atomic D found from our computations [$>$ 0.1 \percc \
for $n$(\MOLH ) = $10^6$ \percc ] is of the order of that
required to explain the high fractionation of deuterium in methanol
and formaldehyde on grain surfaces (Caselli et al. 2002c).

\section{Discussion}
\subsection{Grain size distribution in depleted cores}

The grain size distribution in prestellar cores is not known. It depends on
the timescales for grain coagulation and mantle accretion, relative to the
dynamical
and chemical timescales. In view of this uncertainty, we compared the
results of computations assuming
a unique grain radius with the corresponding calculations assuming a MRN
power--law size distribution; we retained the same value of the grain
surface area per H-nucleus, $n_{g}\,\sigma_{g}/n_{\rm H}$, in both cases.
In calculations with a unique value of $a_g$,
 $n_{g}\,\sigma_{g}/n_{\rm H}
=
1.1\, 10^{-21}\, (0.10/a_{g}$(\mic)) cm$^2$.
 When a power law dependence
$a_{g}^{-r}$ is adopted, providing the exponent $r > 3$ (MRN derived $r =
3.5$),
the lower limit to the radius essentially determines the grain surface
area. When $r =
3.5$, the lower limit to $a_{g}$ is 600 \AA \ for $n_{g}\,\sigma_{g}/n_{\rm H}
=
1.1\, 10^{-21}$ cm$^2$.
We found that the results of calculations with the same value of grain
surface area per H-nucleus were quantitatively similar for the cases of a
unique grain radius and a MRN power--law size distribution.

As shown above, the detection of \HTDP \ in L~1544 by CvTCB
can be explained if $a_{g} <$ 0.1 \mic \, or, equivalently,
if the grain surface area per H-nucleus is
at least   $1.1\, 10^{-21}$ cm$^{2}$, which
is close to the value for grains in diffuse clouds.
This conclusion is consistent
with deductions which have been made from the mm-infrared extinction curve
in prestellar regions: Bianchi et
al. (2003) found that the ratio of mm to NIR dust absorption is
marginally larger in prestellar environments than in diffuse clouds; it is
also consistent with the simulations of Ossenkopf and Henning (1994).

Under conditions of complete heavy element depletion, the thickness of the
grain mantle is expected to
be roughly 40 percent of the grain radius, or 400 \AA \
for 0.1 \mic \ grains. Then,
45 percent of the grain mass is in the refractory core and 55 percent
in the ice mantle. A MRN size distribution yields an ice
mantle
260 \AA \ thick, independent of the grain radius: see Appendix B.

\subsection{Ortho to para ratios}
Our calculations show that the ortho/para
ratios of both \HTHP \ and \HTDP \ are directly dependent on the
\MOLH \ ortho/para ratio.  Thus, measurements of the former would yield
information on the latter.
We emphasize that, whilst steady state may be a reasonable approximation for
most of the chemical species that
we have discussed, this may not be the case for the
\MOLH \ ortho/para ratio.
Thus, reliable measurements of the ortho/para
ratios of either \HTHP \ or \HTDP  could provide information on the
evolution of the source.

Unfortunately, the $1_{01}-0_{00}$ transition to the ground
state of para-\HTDP \ (at 1.37 THz) will be
difficult to excite at the temperatures of prestellar cores and
may be observable only in absorption towards a
background continuum source. On the other hand, \HTHP \ has
no allowed submm transitions and is  best observed in the NIR
at 3.5 \mic \, in absorption towards background sources
(McCall et al. 1999).
It has been assumed that the \HTHP \ which has been observed exists
in regions whose composition is similar to that of
``normal'' interstellar molecular clouds and, furthermore, that
the ortho/para forms are in LTE. In our view, both of these
assumptions may prove to be invalid.
Observations of \HTDP \ and the other members of the
family may help to shed light on this issue.

\subsection{Observations of completely depleted cores}
Observations of pre--protostellar cores, in which heavy elements have
frozen on to grains, are important because they
provide us with information on the structure and
kinematics of the core at the time of its initial gravitational collapse.
 It is important to observe the mm
continuum emission of these objects with high angular resolution
in order to derive the density and temperature distributions for
comparison with theoretical determinations (e.g. Galli et al. 2003).
Likewise, kinematical information may be obtained
from higher angular resolution studies of the $1_{10}-1_{11}$
line of \HTDP .
The present study shows that it is plausible to conclude that
the ortho-\HTDP \ emission detected by CvTCB towards L~1544
arises in completely depleted material. Indeed, these authors
failed by a factor of 10 to account for their observations of L~1544
using estimates based on less extreme assumptions regarding the depletion
of heavy elements from the gas phase.

 Our results can be compared with those of Roberts et al. (2003) who
carried out a time dependent calculation of the chemistry in a prestellar
core with density $3\, 10^6$ \percc  \, including freeze--out. They
found, as do we, that \DTHP \ can be the most abundant ion, though
their results suggest that \HTHP \ becomes the major ion  when all
heavy species freeze out; this may be attributable to a different
way of dealing with grain neutralization of ions.

Further progress in this  area will depend on our capacity to sample
transitions other than the the $1_{10}-1_{11}$
 line of \HTDP . \HTHP \ and
\DTHP \ do not have permanent dipole moments, and hence their detection is
feasible only in absorption in the near infrared.
Our calculations suggest that searches for \DTHP \ should be directed
towards regions
where high heavy element depletion is believed
to have occurred. Searches for
emission in low--lying transitions of \DTWHP \ should also be made: the
abundance of \DTWHP \ in conditions of complete depletion is predicted to
be comparable to that of \HTDP . Under conditions of LTE at temperature
$T$, the relative populations of the lowest ortho ($0_{00}$) and para
($1_{01}$) levels of \DTWHP \ would be

\begin{displaymath}
\frac{n(1_{01})}{n(0_{00})}\, = \, (9/6){\rm exp}(-50.2/T)
\end{displaymath}
where the nuclear spin statistical weights (2$I$ + 1) of the ortho ($I$ =
0, 2) and the para ($I$ = 1) levels are 6 and 3 and their rotational
degeneracies (2$J$ + 1) are 1 and 3, respectively, yielding total
statistical weights of 6 and 9 for $0_{00}$ and $1_{01}$, respectively.
However, under the conditions in prestellar cores, the para/ortho \DTWHP \
ratio (like the ortho/para ratios discussed above, and for related reasons)
is probably much in excess of its LTE value; we estimate that the para and
ortho forms of this species may have comparable abundances. Thus a deep
search for the 692 GHz $1_{10}-1_{01}$
transition of \DTWHP \ seems worthwhile.

\section{Conclusions}
 We summarize here our most important conclusions.
\begin{itemize}
\item We believe that our data
strongly suggest that the detection of \HTDP \ by CvTCB is due
to complete freeze--out of the heavy elements in the core of
L~1544. If so, {\it only} species such as
\HTHP \ and its deuterated counterparts are useful tracers of the
kinematics in the dense centre of L~1544 and other similar objects.
\item Knowledge of the mean grain surface area is needed in
order to understand phenomena such as deuterium fractionation;
it seems likely that this remains true even if some of the
heavy species are in the gas phase.
Our results imply that, in L~1544, the mean grain surface
area per H--atom is above $10^{-21}$ cm$^{2}$, which is comparable
to values found in the diffuse interstellar medium.
\item The degree of ionization and the timescale for ambipolar diffusion,
determined under conditions of complete heavy element depletion, are not
significantly different from values reported elsewhere in the literature,
evaluated assuming degrees of depletion more typical of dense interstellar
clouds. However, we estimate that, under the conditions considered
here, the effect of the coupling between the neutral gas and charged grains
is to increase the ambipolar diffusion timescale by between one and two
orders of magnitude; this finding will be the subject of further
investigation.
\end{itemize}

\appendix

\section{Chemical reactions and rate coefficients}

In this Appendix are specified the reactions included and the rate
coefficients adopted in our chemical model.

Protons have nuclear spin $I$ = 1/2 and are fermions. The ortho and para
forms of H$_2$, H$_{2}^{+}$, H$_{3}^{+}$ and H$_{2}$D$^{+}$ have been
treated as distinct species. Ortho- and para-H$_2$ were assumed to form
hot, on grains, in the ratio 3:1 of their nuclear spin statistical weights.
Subsequent proton--exchange reactions interconvert the ortho and para
forms. Such reactions were assumed to involve only the rotational ground
states of the ortho and para modifications, and the corresponding rate
coefficients take account of the relative statistical weights
(2$I$+1)(2$J$+1) of the products and reactants.

Deuteration of H$_{3}^{+}$ in reactions with HD may be viewed as proton
transfer (from H$_{3}^{+}$ to HD) or deuteron-proton exchange; we allowed
for both viewpoints in our reaction set. Rate coefficients for dissociative
recombination, with electrons in the gas phase or on negatively charged
grains, were assumed to be the same for the ortho and para forms of the
recombining ion.

We note that deuterons have nuclear spin $I$ = 1 and are bosons. Multiply
deuterated species have ortho and para forms. In the present study, we have
not attempted to allow for the consequences of Bose-Einstein statistics
when evaluating the rates of reactions involving multiply--deuterated
species.

\onecolumn
\begin{longtable}[h]
{lccc}
\hline
\hline
 Reaction & $\gamma$ & $\alpha$ & $\beta$\\
\hline
\endhead
\endfoot

 H         +  H      $\rightarrow$ H$_2$(p)                    & 0.25&&\\
 H         +  H      $\rightarrow$ H$_2$(o)                    & 0.75&&\\
 H         +  D      $\rightarrow$ HD                          & 1.00&&\\
 H         +  crp    $\rightarrow$ H$^+$   +  e$^-$            & 0.46&&\\
 He        +  crp    $\rightarrow$ He$^+$  +  e$^-$            & 0.50&&\\
 H$_2$(p)  +  crp    $\rightarrow$ H$^+$   +  H    +  e$^-$    & 0.04&&\\
 H$_2$(o)  +  crp    $\rightarrow$ H$^+$   +  H    +  e$^-$    & 0.04&&\\
 H$_2$(p)  +  crp    $\rightarrow$ H   +   H                   & 1.50&&\\
 H$_2$(o)  +  crp    $\rightarrow$ H   +   H                   & 1.50&&\\
 H$_2$(p)  +  crp    $\rightarrow$ H$_2$$^+$(p) + e$^-$        & 0.96&&\\
 H$_2$(o)  +  crp    $\rightarrow$ H$_2$$^+$(o) + e$^-$        & 0.96&&\\
{\bf  H$^+$         +  H$_2$(o)  $\rightarrow$ H$^+$         +  H$_2$(p) }
& 2.20(-10) & 0.00 &     0.0\\
{\bf  H$_3$$^+$(p)  +  H$_2$(o)  $\rightarrow$ H$_3$$^+$(o)  +  H$_2$(p) }
& 4.40(-10) & 0.00 &     0.0\\
{\bf  H$_3$$^+$(o)  +  H$_2$(o)  $\rightarrow$ H$_3$$^+$(p)  +  H$_2$(p) }
& 1.10(-10) & 0.00 &     0.0\\
 H$^+$         +  H$_2$(p)  $\rightarrow$ H$^+$         +  H$_2$(o)  &
1.98(-09) & 0.00 &   170.5\\
 H$_3$$^+$(p)  +  H$_2$(p)  $\rightarrow$ H$_3$$^+$(o)  +  H$_2$(o)  &
1.98(-09) & 0.00 &   203.4\\
 H$_3$$^+$(o)  +  H$_2$(p)  $\rightarrow$ H$_3$$^+$(p)  +  H$_2$(o)  &
1.98(-09) & 0.00 &   137.6\\
 H$_2$$^+$(o)  +  H$_2$(o)  $\rightarrow$  H$_3$$^+$(p)  +  H        &
1.05(-09) & 0.00 &     0.0\\
 H$_2$$^+$(o)  +  H$_2$(o)  $\rightarrow$  H$_3$$^+$(o)  +  H        &
1.05(-09) & 0.00 &     0.0\\
 H$_2$$^+$(p)  +  H$_2$(o)  $\rightarrow$  H$_3$$^+$(p)  +  H        &
1.05(-09) & 0.00 &     0.0\\
 H$_2$$^+$(p)  +  H$_2$(o)  $\rightarrow$  H$_3$$^+$(o)  +  H        &
1.05(-09) & 0.00 &     0.0\\
 H$_2$$^+$(o)  +  H$_2$(p)  $\rightarrow$  H$_3$$^+$(p)  +  H        &
1.05(-09) & 0.00 &     0.0\\
 H$_2$$^+$(o)  +  H$_2$(p)  $\rightarrow$  H$_3$$^+$(o)  +  H        &
1.05(-09) & 0.00 &     0.0\\
 H$_2$$^+$(p)  +  H$_2$(p)  $\rightarrow$  H$_3$$^+$(p)  +  H        &
2.10(-09) & 0.00 &     0.0\\
 He$^+$    +  H$_2$(p)  $\rightarrow$  H$^+$  +   H    +   He        &
1.10(-13) & -0.24 &    0.0\\
 He$^+$    +  H$_2$(o)  $\rightarrow$  H$^+$  +   H    +   He        &
1.10(-13) & -0.24 &    0.0\\
 H$^+$        +  e$^-$  $\rightarrow$ H    +   photon                &
3.61(-12) & -0.75 &    0.0\\
 H$_2$$^+$(p) +  e$^-$  $\rightarrow$ H    +   H                     &
1.60(-08) & -0.43 &    0.0\\
 H$_2$$^+$(o) +  e$^-$  $\rightarrow$ H    +   H                     &
1.60(-08) & -0.43 &    0.0\\
 He$^+$       +  e$^-$  $\rightarrow$ He   +   photon                &
4.50(-12) & -0.67 &    0.0\\
{\bf  H$_3$$^+$(p) +  e$^-$  $\rightarrow$ H    +   H   +    H  }    &
5.10(-08) & -0.52 &    0.0\\
{\bf  H$_3$$^+$(o) +  e$^-$  $\rightarrow$ H    +   H   +    H  }    &
5.10(-08) & -0.52 &    0.0\\
{\bf  H$_3$$^+$(p) +  e$^-$  $\rightarrow$ H$_2$(p)  +  H       }    &
1.13(-08) & -0.52 &    0.0\\
{\bf  H$_3$$^+$(p) +  e$^-$  $\rightarrow$ H$_2$(o)  +  H       }    &
0.57(-08) & -0.52 &    0.0\\
{\bf  H$_3$$^+$(o) +  e$^-$  $\rightarrow$ H$_2$(o)  +  H       }    &
1.70(-08) & -0.52 &    0.0\\
 D      +  crp   $\rightarrow$  D$^+$   +  e$^-$             & 0.46&&\\
 HD     +  crp   $\rightarrow$  H$^+$   +  D   +  e$^-$      & 0.02&&\\
 HD     +  crp   $\rightarrow$  D$^+$   +  H   +  e$^-$      & 0.02&&\\
 HD     +  crp   $\rightarrow$  H       +  D                 & 1.50&&\\
 HD     +  crp   $\rightarrow$  HD$^+$  +  e$^-$             & 0.96&&\\
 D$^+$  +    H        $\rightarrow$ H$^+$  +  D              & 1.00(-09) &
0.00 &     0.0\\
 H$^+$  +    D        $\rightarrow$ D$^+$  +  H              & 1.00(-09) &
0.00 &    41.0\\
 D$^+$  +    H$_2$(p) $\rightarrow$ H$^+$  +  HD             & 2.10(-09) &
0.00 &     0.0\\
 D$^+$  +    H$_2$(o) $\rightarrow$ H$^+$  +  HD             & 2.10(-09) &
0.00 &     0.0\\
 H$^+$  +    HD       $\rightarrow$ D$^+$  +  H$_2$(p)       & 1.00(-09) &
0.00 &   464.0\\
 H$^+$  +    HD       $\rightarrow$ D$^+$  +  H$_2$(o)       & 1.00(-09) &
0.00 &   634.5\\
 HD$^+$  +   H$_2$(p)  $\rightarrow$ H$_2$D$^+$(p)  +  H     & 5.25(-10) &
0.00 &     0.0\\
 HD$^+$  +   H$_2$(p)  $\rightarrow$ H$_2$D$^+$(o)  +  H     & 5.25(-10) &
0.00 &     0.0\\
 HD$^+$  +   H$_2$(o)  $\rightarrow$ H$_2$D$^+$(p)  +  H     & 5.25(-10) &
0.00 &     0.0\\
 HD$^+$  +   H$_2$(o)  $\rightarrow$ H$_2$D$^+$(o)  +  H     & 5.25(-10) &
0.00 &     0.0\\
 HD$^+$  +   H$_2$(p)  $\rightarrow$ H$_3$$^+$(p)   +  D     & 1.05(-09) &
0.00 &     0.0\\
 HD$^+$  +   H$_2$(o)  $\rightarrow$ H$_3$$^+$(o)   +  D     & 5.25(-09) &
0.00 &     0.0\\
 HD$^+$  +   H$_2$(o)  $\rightarrow$ H$_3$$^+$(p)   +  D     & 5.25(-09) &
0.00 &     0.0\\
 H$_2$$^+$(p)  +  HD   $\rightarrow$   H$_2$D$^+$(p)  +  H               &
5.25(-10) & 0.00 &     0.0\\
 H$_2$$^+$(p)  +  HD   $\rightarrow$   H$_2$D$^+$(o)  +  H               &
5.25(-10) & 0.00 &     0.0\\
 H$_2$$^+$(o)  +  HD   $\rightarrow$   H$_2$D$^+$(p)  +  H               &
5.25(-10) & 0.00 &     0.0\\
 H$_2$$^+$(o)  +  HD   $\rightarrow$   H$_2$D$^+$(o)  +  H               &
5.25(-10) & 0.00 &     0.0\\
 H$_2$$^+$(p)  +  HD   $\rightarrow$   H$_3$$^+$(p)  +  D                &
1.05(-09) & 0.00 &     0.0\\
 H$_2$$^+$(o)  +  HD   $\rightarrow$   H$_3$$^+$(p)  +  D                &
5.25(-10) & 0.00 &     0.0\\
 H$_2$$^+$(o)  +  HD   $\rightarrow$   H$_3$$^+$(o)  +  D                &
5.25(-10) & 0.00 &     0.0\\
 H$_3$$^+$(p)  +  D    $\rightarrow$   H$_2$D$^+$(p)  +  H               &
0.66(-09) & 0.00 &     0.0\\
 H$_3$$^+$(p)  +  D    $\rightarrow$   H$_2$D$^+$(o)  +  H               &
0.33(-09) & 0.00 &     0.0\\
 H$_3$$^+$(o)  +  D    $\rightarrow$   H$_2$D$^+$(o)  +  H               &
1.00(-09) & 0.00 &     0.0\\
 H$_2$D$^+$(p)  +  H   $\rightarrow$    H$_3$$^+$(p)  +  D               &
1.00(-09) & 0.00 &   632.0\\
 H$_2$D$^+$(o)  +  H   $\rightarrow$    H$_3$$^+$(o)  +  D               &
0.50(-09) & 0.00 &   578.4\\
 H$_2$D$^+$(o)  +  H   $\rightarrow$    H$_3$$^+$(p)  +  D               &
0.50(-09) & 0.00 &   545.5\\
{\bf  H$_3$$^+$(p)  +  HD   $\rightarrow$   H$_2$D$^+$(p)  +  H$_2$(p) } &
1.17(-10) & 0.00 &     0.0\\
{\bf  H$_3$$^+$(p)  +  HD   $\rightarrow$   H$_2$D$^+$(p)  +  H$_2$(o) } &
1.17(-10) & 0.00 &     0.0\\
{\bf  H$_3$$^+$(p)  +  HD   $\rightarrow$   H$_2$D$^+$(o)  +  H$_2$(p) } &
5.83(-11) & 0.00 &     0.0\\
{\bf  H$_3$$^+$(p)  +  HD   $\rightarrow$   H$_2$D$^+$(o)  +  H$_2$(o) } &
5.83(-11) & 0.00 &    25.0\\
{\bf  H$_3$$^+$(o)  +  HD   $\rightarrow$   H$_2$D$^+$(p)  +  H$_2$(o) } &
1.75(-10) & 0.00 &     0.0\\
{\bf  H$_3$$^+$(o)  +  HD   $\rightarrow$   H$_2$D$^+$(o)  +  H$_2$(o) } &
1.75(-10) & 0.00 &     0.0\\
{\bf  H$_2$D$^+$(p)  +  H$_2$(o)  $\rightarrow$  H$_2$D$^+$(o)  +  H$_2$(p)
} & 1.98(-09) & 0.00 &     0.0\\
{\bf  H$_2$D$^+$(o)  +  H$_2$(p)  $\rightarrow$  H$_2$D$^+$(p)  +  H$_2$(o)
} & 1.98(-09) & 0.00 &    84.0\\
{\bf  H$_2$D$^+$(o)  +  H$_2$(o)  $\rightarrow$  H$_2$D$^+$(p)  +  H$_2$(p)
} & 2.44(-11) & 0.00 &     0.0\\
 H$_2$D$^+$(p)  +  H$_2$(p)  $\rightarrow$  H$_2$D$^+$(o)  +  H$_2$(o)   &
1.98(-09) & 0.00 &   257.0\\
 H$_2$D$^+$(p)  +  H$_2$(p)  $\rightarrow$  H$_3$$^+$(p)  +  HD          &
1.40(-10) & 0.00 &   232.0\\
 H$_2$D$^+$(p)  +  H$_2$(o)  $\rightarrow$  H$_3$$^+$(p)  +  HD          &
0.70(-10) & 0.00 &    61.5\\
 H$_2$D$^+$(p)  +  H$_2$(o)  $\rightarrow$  H$_3$$^+$(o)  +  HD          &
0.70(-10) & 0.00 &    94.4\\
 H$_2$D$^+$(o)  +  H$_2$(p)  $\rightarrow$  H$_3$$^+$(p)  +  HD          &
1.40(-10) & 0.00 &   145.5\\
{\bf  H$_2$D$^+$(o)  +  H$_2$(o)  $\rightarrow$  H$_3$$^+$(p)  +  HD }   &
0.70(-10) & 0.00 &     0.0\\
{\bf  H$_2$D$^+$(o)  +  H$_2$(o)  $\rightarrow$  H$_3$$^+$(o)  +  HD }   &
0.70(-10) & 0.00 &     8.0\\
 He$^+$   +  HD   $\rightarrow$   H$^+$  +  D  +  He                     &
5.50(-14) & -0.24 &    0.0\\
 He$^+$   +  HD   $\rightarrow$   D$^+$  +  H  +  He                     &
5.50(-14) & -0.24 &    0.0\\
 D$^+$    +  e$^-$  $\rightarrow$  D   +  photon                 &
3.61(-12) & -0.75 &     0.0\\
 HD$^+$   +  e$^-$  $\rightarrow$  H   +  D                      &
3.40(-09) & -0.40 &     0.0\\
{\bf  H$_2$D$^+$(p)   +  e$^-$  $\rightarrow$  H   +  H   +  D } &
4.96(-08) & -0.52 &     0.0\\
{\bf  H$_2$D$^+$(o)   +  e$^-$  $\rightarrow$  H   +  H   +  D } &
4.96(-08) & -0.52 &     0.0\\
{\bf  H$_2$D$^+$(p)   +  e$^-$  $\rightarrow$  HD   +  H       } &
1.36(-08) & -0.52 &     0.0\\
{\bf  H$_2$D$^+$(o)   +  e$^-$  $\rightarrow$  HD   +  H       } &
1.36(-08) & -0.52 &     0.0\\
{\bf  H$_2$D$^+$(p)   +  e$^-$  $\rightarrow$  H$_2$(p)   +  D } &
4.76(-09) & -0.52 &     0.0\\
{\bf  H$_2$D$^+$(o)   +  e$^-$  $\rightarrow$  H$_2$(o)   +  D } &
4.76(-09) & -0.52 &     0.0\\
 D$^+$   +  H  $\rightarrow$  HD$^+$   +  photon                & 3.90(-19)
& 1.80  &     0.0\\
 H$^+$   +  D  $\rightarrow$  HD$^+$   +  photon                & 3.90(-19)
& 1.80  &     0.0\\
 H$_2$$^+$(p)   +  D  $\rightarrow$  H$_2$D$^+$(p)   +  photon  & 7.00(-18)
& 1.80  &     0.0\\
 HD$^+$   +  H  $\rightarrow$  H$_2$D$^+$(p)   +  photon        & 1.20(-17)
& 1.80  &     0.0\\
 HD$^+$   +  H  $\rightarrow$  H$^+$   +  HD                    & 6.40(-10)
& 0.00  &     0.0\\
 H$_2$$^+$(p)   +  D  $\rightarrow$  D$^+$   +  H$_2$(p)        & 6.40(-10)
& 0.00  &     0.0\\
 HD$^+$   +  H  $\rightarrow$  H$_2$$^+$(p)   +  D              & 1.00(-09)
& 0.00  &   154.0\\
 H$_2$$^+$(p)   +  D  $\rightarrow$  HD$^+$   +  H              & 1.00(-09)
& 0.00  &     0.0\\
 D$_2$  +  crp  $\rightarrow$  D$^+$  +  D  +  e$^-$            & 0.04&&\\
 D$_2$  +  crp  $\rightarrow$  D  +  D                          & 1.50&&\\
 D$_2$  +  crp  $\rightarrow$  D$_2$$^+$  +  e$^-$              & 0.96&&\\
 H$^+$  +  D$_2$  $\rightarrow$  D$^+$  +  HD                    &
2.10(-09) & 0.00 &   491.0\\
 D$^+$  +  HD  $\rightarrow$  H$^+$  +  D$_2$                    &
1.00(-09) & 0.00 &     0.0\\
 D$_2$$^+$  +  H$_2$(p)  $\rightarrow$  H$_2$D$^+$(p)  +  D      &
1.05(-09) & 0.00 &     0.0\\
 D$_2$$^+$  +  H$_2$(o)  $\rightarrow$  H$_2$D$^+$(o)  +  D      &
1.05(-09) & 0.00 &     0.0\\
 D$_2$$^+$  +  H$_2$(p)  $\rightarrow$  HD$_2$$^+$  +  H         &
1.05(-09) & 0.00 &     0.0\\
 D$_2$$^+$  +  H$_2$(o)  $\rightarrow$  HD$_2$$^+$  +  H         &
1.05(-09) & 0.00 &     0.0\\
 H$_2$$^+$(p)  +  D$_2$  $\rightarrow$  H$_2$D$^+$(p)  +  D      &
1.05(-09) & 0.00 &     0.0\\
 H$_2$$^+$(o)  +  D$_2$  $\rightarrow$  H$_2$D$^+$(o)  +  D      &
1.05(-09) & 0.00 &     0.0\\
 H$_2$$^+$(p)  +  D$_2$  $\rightarrow$  HD$_2$$^+$  +  H         &
1.05(-09) & 0.00 &     0.0\\
 H$_2$$^+$(o)  +  D$_2$  $\rightarrow$  HD$_2$$^+$  +  H         &
1.05(-09) & 0.00 &     0.0\\
 HD$^+$  +  HD  $\rightarrow$  H$_2$D$^+$(p)  +  D               &
5.25(-10) & 0.00 &     0.0\\
 HD$^+$  +  HD  $\rightarrow$  H$_2$D$^+$(o)  +  D               &
5.25(-10) & 0.00 &     0.0\\
 HD$^+$  +  HD  $\rightarrow$  HD$_2$$^+$  +  H                  &
1.05(-09) & 0.00 &     0.0\\
 H$_2$D$^+$(p)  +  D  $\rightarrow$  HD$_2$$^+$  +  H            &
1.00(-09) & 0.00 &     0.0\\
 H$_2$D$^+$(o)  +  D  $\rightarrow$  HD$_2$$^+$  +  H            &
1.00(-09) & 0.00 &     0.0\\
 HD$_2$$^+$  +  H  $\rightarrow$  H$_2$D$^+$(p)  +  D            &
1.00(-09) & 0.00 &   600.0\\
 HD$_2$$^+$  +  H   $\rightarrow$  H$_2$D$^+$(o)  +  D           &
1.00(-09) & 0.00 &   513.5\\
 H$_3$$^+$(o)  +  D$_2$  $\rightarrow$  H$_2$D$^+$(o)  +  HD     &
3.50(-10) & 0.00 &     0.0\\
 H$_3$$^+$(o)  +  D$_2$  $\rightarrow$  HD$_2$$^+$  +  H$_2$(o)  &
3.50(-10) & 0.00 &     0.0\\
 H$_3$$^+$(p)  +  D$_2$  $\rightarrow$  HD$_2$$^+$  +  H$_2$(p)  &
2.33(-10) & 0.00 &     0.0\\
 H$_3$$^+$(p)  +  D$_2$  $\rightarrow$  HD$_2$$^+$  +  H$_2$(o)  &
1.17(-10) & 0.00 &     0.0\\
 H$_3$$^+$(p)  +  D$_2$  $\rightarrow$  H$_2$D$^+$(p)  +  HD     &
2.33(-10) & 0.00 &     0.0\\
 H$_3$$^+$(p)  +  D$_2$  $\rightarrow$  H$_2$D$^+$(o)  +  HD     &
1.17(-10) & 0.00 &     0.0\\
{\bf  H$_2$D$^+$(p)  +  HD  $\rightarrow$  HD$_2$$^+$  +  H$_2$(p) } &
2.60(-10) & 0.00 &     0.0\\
{\bf  H$_2$D$^+$(o)  +  HD  $\rightarrow$  HD$_2$$^+$  +  H$_2$(o) } &
2.60(-10) & 0.00 &     0.0\\
 H$_2$D$^+$(o)  +  HD  $\rightarrow$  H$_3$$^+$(o)  +  D$_2$     &
1.30(-10) & 0.00 &   108.0\\
 H$_2$D$^+$(o)  +  HD  $\rightarrow$  H$_3$$^+$(p)  +  D$_2$     &
1.30(-10) & 0.00 &    75.0\\
 H$_2$D$^+$(p)  +  HD  $\rightarrow$  H$_3$$^+$(p)  +  D$_2$     &
2.60(-10) & 0.00 &   161.5\\
 HD$_2$$^+$  +  H$_2$(p)  $\rightarrow$  H$_2$D$^+$(p)  +  HD    &
1.00(-10) & 0.00 &   190.0\\
 HD$_2$$^+$  +  H$_2$(p)  $\rightarrow$  H$_2$D$^+$(o)  +  HD    &
1.00(-10) & 0.00 &   277.0\\
 HD$_2$$^+$  +  H$_2$(o)  $\rightarrow$  H$_2$D$^+$(o)  +  HD    &
1.00(-10) & 0.00 &   106.0\\
{\bf  HD$_2$$^+$  +  H$_2$(o)  $\rightarrow$  H$_2$D$^+$(p)  +  HD } &
1.00(-10) & 0.00 &    20.1\\
 HD$_2$$^+$  +  H$_2$(p)  $\rightarrow$  H$_3$$^+$(p)  +  D$_2$  &
2.00(-10) & 0.00 &   320.0\\
 HD$_2$$^+$  +  H$_2$(o)  $\rightarrow$  H$_3$$^+$(p)  +  D$_2$  &
1.00(-10) & 0.00 &   149.5\\
 HD$_2$$^+$  +  H$_2$(o)  $\rightarrow$  H$_3$$^+$(o)  +  D$_2$  &
1.00(-10) & 0.00 &   182.4\\
 He$^+$  +  D$_2$  $\rightarrow$  D$^+$  +   D  +  He            &
1.10(-13) & -0.24 &    0.0\\
 He$^+$  +  D$_2$  $\rightarrow$  D$_2$$^+$  +  He               &
2.50(-14) & 0.00 &     0.0\\
 D$_2$$^+$  +  e$^-$  $\rightarrow$  D  +  D                     &
3.40(-09) & -0.40 &    0.0\\
{\bf  HD$_2$$^+$  +  e$^-$  $\rightarrow$  D  +  D  +  H }       &
4.96(-08) & -0.52 &    0.0\\
{\bf  HD$_2$$^+$  +  e$^-$  $\rightarrow$  HD  +  D      }       &
1.36(-08) & -0.52 &    0.0\\
{\bf  HD$_2$$^+$  +  e$^-$  $\rightarrow$  D$_2$  +  H   }       &
4.76(-09) & -0.52 &    0.0\\
 D$_2$$^+$  +  H  $\rightarrow$  H$^+$  +  D$_2$                 &
6.40(-10) & 0.00 &    0.0\\
 HD$^+$  +  D  $\rightarrow$  D$^+$  +  HD                       &
6.40(-10) & 0.00 &    0.0\\
 HD$^+$  +  D  $\rightarrow$  D$_2$$^+$  +  H                    &
1.00(-09) & 0.00 &    0.0\\
 D$_2$$^+$  +  H  $\rightarrow$  HD$^+$  +  D                    &
1.00(-09) & 0.00 &  472.0\\
 D$_2$$^+$  +  D  $\rightarrow$  D$^+$  +  D$_2$                 &
6.40(-10) & 0.00 &     0.0\\
 D$_2$$^+$  +  HD  $\rightarrow$  HD$_2$$^+$  +  D               &
1.05(-09) & 0.00 &     0.0\\
 D$_2$$^+$  +  HD  $\rightarrow$  D$_3$$^+$  +  H                &
1.05(-09) & 0.00 &     0.0\\
 HD$^+$  +  D$_2$  $\rightarrow$  HD$_2$$^+$  +  D               &
1.05(-09) & 0.00 &     0.0\\
 HD$^+$  +  D$_2$  $\rightarrow$  D$_3$$^+$  +  H                &
1.05(-09) & 0.00 &     0.0\\
 HD$_2$$^+$  +  D  $\rightarrow$  D$_3$$^+$  +  H                &
1.00(-09) & 0.00 &     0.0\\
 D$_3$$^+$  +  H  $\rightarrow$  HD$_2$$^+$  +  D                &
1.00(-09) & 0.00 &   655.0\\
{\bf  HD$_2$$^+$  +  HD  $\rightarrow$  D$_3$$^+$  +  H$_2$(p) } &
1.00(-10) & 0.00 &     0.0\\
{\bf  HD$_2$$^+$  +  HD  $\rightarrow$  D$_3$$^+$  +  H$_2$(o) } &
1.00(-10) & 0.00 &     0.0\\
 HD$_2$$^+$  +  HD  $\rightarrow$  H$_2$D$^+$(p)  +  D$_2$       &
1.00(-10) & 0.00 &   110.0\\
 HD$_2$$^+$  +  HD  $\rightarrow$  H$_2$D$^+$(o)  +  D$_2$       &
1.00(-10) & 0.00 &   196.5\\
 H$_2$D$^+$(p)  +  D$_2$  $\rightarrow$  HD$_2$$^+$  +  HD       &
8.50(-10) & 0.00 &     0.0\\
 H$_2$D$^+$(o)  +  D$_2$  $\rightarrow$  HD$_2$$^+$  +  HD       &
8.50(-10) & 0.00 &     0.0\\
 H$_2$D$^+$(p)  +  D$_2$  $\rightarrow$  D$_3$$^+$  +  H$_2$(p)  &
8.50(-10) & 0.00 &     0.0\\
 H$_2$D$^+$(o)  +  D$_2$  $\rightarrow$  D$_3$$^+$  +  H$_2$(o)  &
8.50(-10) & 0.00 &     0.0\\
 D$_3$$^+$  +  H$_2$(p)  $\rightarrow$  H$_2$D$^+$(p)  +  D$_2$  &
1.50(-09) & 0.00 &   352.0\\
 D$_3$$^+$  +  H$_2$(o)  $\rightarrow$  H$_2$D$^+$(o)  +  D$_2$  &
1.50(-09) & 0.00 &   268.0\\
 D$_3$$^+$  +  H$_2$(p)  $\rightarrow$  HD$_2$$^+$  +  HD        &
1.50(-09) & 0.00 &   241.0\\
 D$_3$$^+$  +  H$_2$(o)  $\rightarrow$  HD$_2$$^+$  +  HD        &
1.50(-09) & 0.00 &    70.5\\
 D$_3$$^+$  +  e$^-$  $\rightarrow$  D  +  D  +  D               &
2.03(-08) & -0.50 &    0.0\\
 D$_3$$^+$  +  e$^-$  $\rightarrow$  D$_2$  +  D                 &
6.75(-09) & -0.50 &    0.0\\
 D$_2$$^+$  +  D$_2$  $\rightarrow$  D$_3$$^+$  +  D             &
2.10(-09) & 0.00 &     0.0\\
 D$_3$$^+$  +  HD  $\rightarrow$  HD$_2$$^+$  +  D$_2$           &
1.50(-09) & 0.00 &   161.5\\
 g$^0$  +  secpho  $\rightarrow$  g$^+$  +  e$^-$                 &
0.63(08)&&\\
 g$^-$  +  secpho  $\rightarrow$  g$^0$  +  e$^-$                 &
0.41(09)&&\\
 g$^0$  +  e$^-$  $\rightarrow$  g$^-$  +  photon                 &
6.90(-05) & 0.50 &     0.0\\
{\bf  g$^-$  +  H$^+$  $\rightarrow$  g$^0$  +  H                     } &
1.60(-06) & 0.50 &  0.0\\
{\bf  g$^-$  +  H$_3$$^+$(p)  $\rightarrow$  g$^0$  +  H$_2$(p)  +  H } &
3.07(-07) & 0.50 &  0.0\\
{\bf  g$^-$  +  H$_3$$^+$(p)  $\rightarrow$  g$^0$  +  H$_2$(o)  +  H } &
1.54(-07) & 0.50 &  0.0\\
{\bf  g$^-$  +  H$_3$$^+$(o)  $\rightarrow$  g$^0$  +  H$_2$(o)  +  H } &
4.61(-07) & 0.50 &  0.0\\
{\bf  g$^-$  +  H$_3$$^+$(p)  $\rightarrow$  g$^0$  +  3H             } &
4.61(-07) & 0.50 &  0.0\\
{\bf  g$^-$  +  H$_3$$^+$(o)  $\rightarrow$  g$^0$  +  3H             } &
4.61(-07) & 0.50 &  0.0\\
 g$^-$  +  He$^+$  $\rightarrow$  g$^0$  +  He                    &
8.00(-07) & 0.50 &     0.0\\
 g$^0$  +  H$^+$  $\rightarrow$  g$^+$  +  H                      &
1.60(-06) & 0.50 &     0.0\\
 g$^0$  +  H$_3$$^+$(p)  $\rightarrow$  g$^+$  +  H$_2$(p)  +  H  &
3.07(-07) & 0.50 &     0.0\\
 g$^0$  +  H$_3$$^+$(p)  $\rightarrow$  g$^+$  +  H$_2$(o)   H    &
1.54(-07) & 0.50 &     0.0\\
 g$^0$  +  H$_3$$^+$(o)  $\rightarrow$  g$^+$  +  H$_2$(o)   H    &
4.61(-07) & 0.50 &     0.0\\
 g$^0$  +  H$_3$$^+$(p)  $\rightarrow$  g$^+$  +  3H              &
4.61(-07) & 0.50 &     0.0\\
 g$^0$  +  H$_3$$^+$(o)  $\rightarrow$  g$^+$  +  3H              &
4.61(-07) & 0.50 &     0.0\\
 g$^0$  +  He$^+$  $\rightarrow$  g$^+$  +  He                    &
8.00(-07) & 0.50 &     0.0\\
 g$^+$  +  e$^-$  $\rightarrow$  g$^0$  +  photon                 &
6.90(-05) & 0.50 &     0.0\\
 g$^-$  +  D$^+$  $\rightarrow$  g$^0$  +  D                      &
1.13(-06) & 0.50 &     0.0\\
{\bf  g$^-$  +  H$_2$D$^+$(p)  $\rightarrow$  g$^0$  +  H$_2$(p)  +  D } &
1.33(-07) & 0.50 &  0.0\\
{\bf  g$^-$  +  H$_2$D$^+$(p)  $\rightarrow$  g$^0$  +  HD  +  H       } &
2.66(-07) & 0.50 &  0.0\\
{\bf  g$^-$  +  H$_2$D$^+$(o)  $\rightarrow$  g$^0$  +  H$_2$(o)  +  D } &
1.33(-07) & 0.50 &  0.0\\
{\bf  g$^-$  +  H$_2$D$^+$(o)  $\rightarrow$  g$^0$  +  HD  +  H       } &
2.66(-07) & 0.50 &  0.0\\
{\bf  g$^-$  +  H$_2$D$^+$(p)  $\rightarrow$  g$^0$  +  D  +  2H       } &
3.99(-07) & 0.50 &  0.0\\
{\bf  g$^-$  +  H$_2$D$^+$(o)  $\rightarrow$  g$^0$  +  D  +  2H       } &
3.99(-07) & 0.50 &  0.0\\
 g$^0$  +  D$^+$  $\rightarrow$  g$^+$  +  D                      &
1.13(-06) & 0.50 &     0.0\\
 g$^0$  +  H$_2$D$^+$(p)  $\rightarrow$  g$^+$  +  H$_2$(p)  +  D &
1.33(-07) & 0.50 &     0.0\\
 g$^0$  +  H$_2$D$^+$(p)  $\rightarrow$  g$^+$  +  HD  +  H       &
2.66(-07) & 0.50 &     0.0\\
 g$^0$  +  H$_2$D$^+$(o)  $\rightarrow$  g$^+$  +  H$_2$(o)  +  D &
1.33(-07) & 0.50 &     0.0\\
 g$^0$  +  H$_2$D$^+$(o)  $\rightarrow$  g$^+$  +  HD  +  H       &
2.66(-07) & 0.50 &     0.0\\
 g$^0$  +  H$_2$D$^+$(p)  $\rightarrow$  g$^+$  +  D  +  2H       &
3.99(-07) & 0.50 &     0.0\\
 g$^0$  +  H$_2$D$^+$(o)  $\rightarrow$  g$^+$  +  D  +  2H       &
3.99(-07) & 0.50 &     0.0\\
{\bf  g$^-$  +  HD$_2$$^+$  $\rightarrow$  g$^0$  +  HD  +  D    } &
2.38(-07) & 0.50 &    0.0\\
{\bf  g$^-$  +  HD$_2$$^+$  $\rightarrow$  g$^0$  +  D$_2$  +  H } &
1.19(-07) & 0.50 &    0.0\\
{\bf  g$^-$  +  HD$_2$$^+$  $\rightarrow$  g$^0$  +  2D  +  H    } &
3.57(-07) & 0.50 &    0.0\\
 g$^0$  +  HD$_2$$^+$  $\rightarrow$  g$^+$  +  HD  +  D          &
2.38(-07) & 0.50 &     0.0\\
 g$^0$  +  HD$_2$$^+$  $\rightarrow$  g$^+$  +  D$_2$  +  H       &
1.19(-07) & 0.50 &     0.0\\
 g$^0$  +  HD$_2$$^+$  $\rightarrow$  g$^+$  +  2D  +  H          &
3.57(-07) & 0.50 &     0.0\\
{\bf  g$^-$  +  D$_3$$^+$  $\rightarrow$  g$^0$  +  D$_2$  +  D } &
3.26(-07) & 0.50 &     0.0\\
{\bf  g$^-$  +  D$_3$$^+$  $\rightarrow$  g$^0$  +  3D          } &
3.26(-07) & 0.50 &     0.0\\
 g$^0$  +  D$_3$$^+$  $\rightarrow$  g$^+$  +  D$_2$  +  D        &
3.26(-07) & 0.50 &     0.0\\
 g$^0$  +  D$_3$$^+$  $\rightarrow$  g$^+$  +  3D                 &
3.26(-07) & 0.50 &     0.0\\

\caption{Reactions and rate coefficients adopted in our chemical model.
The parameters $\alpha$, $\beta$, and $\gamma$ define the rate coefficients
$k$ cm$^3$ s$^{-1}$ at temperature $T$ through the relation $k$ =
\~{J}$(a_g, T)\gamma (T/300)^{\alpha } {\rm exp}(-\beta /T)$; \~{J} allows
for Coulomb focusing in reactions of positive ions and negatively charged
grains (Draine and Sutin 1987, equation 3.4); $a_g$ = 0.1 \mic \ is adopted
in the Table. The rates s$^{-1}$ of reactions induced directly by cosmic
rays (crp) or indirectly by the secondary electrons, which generate H$_2$
fluorescence photons (secpho), are given by $\gamma \zeta$, where $\zeta$
is the rate of cosmic ray ionization of H$_2$. In the first three reactions
(formation of H$_2$ or HD on grain surfaces), $\gamma$ denotes the fraction
of such reactions which form the specified product; the rate is calculated
internally in the program from the grain parameters. Key reactions are in
bold face. Numbers in parentheses are powers of 10. }
\end{longtable}

\twocolumn

\section{Ice Mantle Thickness}

 In this Appendix, we derive estimates of the ice--mantle
thickness in completely depleted cores. We follow the approach
of FPdesF.

Following FPdesF, we assume the refractory cores to be composed of
silicates (in practice, olivine, MgFeSiO$_4$) or graphite. The mass of the
cores, per unit volume of gas, is
$\sum n({\rm X}) M $, where the summation extends over the elements X, and
$M$ denotes the elemental mass.
We define $\alpha_{c}$ through the relation $\sum n({\rm X}) M\, =\,
\alpha_{c}\, n_{\rm H} m_{\rm H}$, and $\alpha_{c}\, =\, 7.84\, 10^{-3}$,
which corresponds to a fraction of 0.0056 by mass
relative to the gas--phase H and He.

Assuming that all elements heavier than He are in the grain cores or frozen
in the mantles, the total dust--to--gas mass density ratio is 0.0125, and
the mass per unit volume of the gas of the ice in the mantle, is
$\alpha_{i}\, n_{\rm H}m_{\rm H}$, where $\alpha_{i}\, =\, 9.63 \,
10^{-3}$.

The volume $V_{i}$
of a grain ice mantle is determined by the relation

\begin{equation}
 n_{c}\, V_{i}\, \rho_{i}\, = \, \alpha_{i}\, n_{\rm H} m_{\rm H}
\end{equation}
where $n_{c}$ is the number of grain cores (i.e. the number of grains) per
unit volume of the gas and $\rho_{i}$ is the density of the mantle
material; $n_{c}$ is defined by

\begin{equation}
n_{c}\, V_{c}\, \rho_{c}\, = \, \alpha_{c}\, n_{\rm H} m_{\rm H}
\end{equation}
where $\rho_{c}$ is the density and $V_{c}$ is the volume of refractory
core material. We adopt
$\rho_{c}$ = 3 gm cm$^{-3}$ and $\rho_{i}$ = 1 gm cm$^{-3}$.
Hence, the ratio of mantle to core volume
is
\begin{equation}
  V_{i}/V_{c}\, =\, \frac{\alpha_{i}\rho_{c}}
           {\alpha_{c}\rho_{i}}
\end{equation}

Substituting the above numbers, we find that the ratio
$V_{i}/V_{c}$ = 3.72, or a fraction
0.79 of the grain volume is occupied by the mantle.  The ratio of
mantle thickness to core radius is 0.67, and the ratio of
mantle thickness to grain (core + mantle) radius is 0.40, in the limit of
complete
heavy element depletion. Thus, for a grain of
radius $a_{g}$ = 0.1 \mic , the mantle thickness is
400 \AA.

This analysis can be generalized to the case of a MRN power--law
size distribution, using the formalism of Le Bourlot et al. (1995). If the
lower limit of the size distribution is taken to be $a_{m}$ = 0.01 \mic \
(and the upper limit is $a_{M}$ = 0.3 \mic ), then the thickness $\delta$
of the mantle is obtained by solving the equation

\begin{equation}
f_4(V_{i}/V_{c})\, = \, (f_1\delta ^3 + 3f_2\delta ^2 + 3f_3\delta )
\end{equation}
where $f_n$ is defined by
\begin{equation}
f_n\, = \, (a_{M}^{n-r} - a_{m}^{n-r})/(n-r)
\end{equation}
and $r$ = 3.5 for the MRN grain size distribution. Solving for $\delta$, we
find that the mantle thickness is 260 \AA,  independent of $a$ for the
specified values of $a_{m}$ (and $a_{M}$).
Thus, very small grains are not present under conditions of complete heavy
element depletion, owing to the formation of an ice mantle around the core.

\begin{acknowledgements}
It is a pleasure to thank Evelyne Roueff and Eric Herbst for informative
and helpful discussions, Paola Caselli for her comments on
the text, {\bf and the referee, Ted Bergin, for a constructive and
thought-provoking report}. CMW thanks the Ecole Normale Sup\'{e}rieure in
Paris for hospitality
during a stay when part of the work described here was
carried out as well as the Max Planck Institute in Bonn
for hospitality on diverse occasions.
\end{acknowledgements}

\end{document}